
\input phyzzx

\hoffset=0.2truein
\voffset=0.1truein
\hsize=6truein
\def\TITLEPAGE{\frontpagetrue}
\def\CALT#1{\hbox to\hsize{\tenpoint \baselineskip=12pt
        \hfil\vtop{
        \hbox{\strut CALT-68-#1}}}}

\def\CALTECH{
        \address{California Institute of Technology,
Pasadena, CA 91125}}

\def\AUTHOR#1{\vskip .2in \centerline{#1}}
\def\ANDAUTHOR#1{\smallskip \centerline{\it and} \smallskip
\centerline{#1}}
\def\ABSTRACT#1{\vskip .2in \vfil \centerline{\twelvepoint
\bf Abstract}
        #1 \vfil}
\def\ENDTITLEPAGE{\vfil\eject\pageno=1}

\tolerance=10000
\hfuzz=5pt

\def\vslash{\rlap{/}v}
\TITLEPAGE
\CALT{1919}         
\bigskip
\titlestyle {Excited $\Lambda_Q$ Baryons in the Large $N_c$ Limit\foot{Work
 supported
 in part by the U.S. Dept. of Energy
under Grant no. DE-FG03-92-ER 40701.}}
\AUTHOR{Chi-Keung Chow}
\ANDAUTHOR{Mark B. Wise}
\CALTECH
\ABSTRACT{The spectrum of excited $\Lambda_Q$-type heavy baryons is considered
 in the large $N_c$ limit. The universal form factors for $\Lambda_b$
 semileptonic decay to excited charmed baryons are calculated in the large
$N_c$ limit.  We find that the Bjorken sum rule (for the slope of the
Isgur--Wise function) and Voloshin sum rule (for the mass of the light degrees
of freedom) are saturated by
 the first doublet of excited $\Lambda_Q$ states.}
\ENDTITLEPAGE

\eject

Experimental evidence for a doublet of excited charm baryons has recently been
  obtained [1].  They have masses 340 MeV and 308 MeV above the $\Lambda_c$.
It is
 natural to interpret these states as the spin 3/2 and 1/2 isospin zero members
 of a doublet that has spin parity of the light degrees of freedom,
 $s_\ell^{\pi_\ell} = 1^-$.

Properties of these excited $\Lambda_c$ baryons can be estimated using the
 nonrelativistic constituent quark model [2].  In this phenomenological model
the
 observed excited $\Lambda_c$ baryons have quark content $cud$ with the $ud$
 pair in an isospin zero and spin zero state like the ground state $\Lambda_c$.
  However, unlike the ground state, in these excited $\Lambda_c$ baryons the
$ud$ pair has a unit of orbital angular momentum about the charm quark.

In this paper we use the large $N_c$ limit [3] to derive properties of excited
 $\Lambda_Q$ baryons.  The predictive power of this limit arises because the
 number of the light quarks in these baryons, $N_c -1$, becomes large [4] as
$N_c
 \rightarrow \infty$.  In the physical three color case there are only two
 light quarks in a baryon with a single heavy quark and so we expect the large
 $N_c$ limit to have only qualitative relevance.  Nevertheless, unlike the
 nonrelativistic constituent quark model, the large $N_c$ limit is the leading
 term in a systematic expansion of QCD and because of this we find its
 consequences interesting.

In the large $N_c$ limit the light baryons, $n, p$, $\Delta$, $etc$., can be
 viewed as solitons in the chiral Lagrangian for pion self interactions [5].
  The
 baryons containing  a single heavy charm (or bottom) quark are bound states of
 these solitons with $D$ and $D^*$  (or $B$ and $B^*$) mesons [6-12].  In this
 paper
 we use the bound state soliton picture to derive properties of the excited
 baryons
that contain a heavy quark.  However, since the results we derive do not depend
 on the couplings in the chiral Lagrangian for pion self interactions and
 pion-heavy meson interactions, they are interpreted as predictions of the
large
 $N_c$ limit.  It should be possible to derive these results in other ways
 [13],
 however, the bound state soliton picture of Callan and Klebanov provides a
convenient way to explore the consequences of the $N_c \rightarrow \infty$
 limit for baryons containing a single heavy quark.

For baryons containing a single heavy quark the nucleon mass $M_B$ plays a
 special role.  In the large $N_c$ limit the mass of the light degrees of
freedom, $\bar \Lambda$, is equal to $M_B$.  The
equality, $\bar\Lambda \equiv M_{\Lambda_Q}-m_Q = M_B$, has a simple physical
 origin.  In the nucleon a light quark responds to the mean color field created
 by $N_c-1$ other quarks.  With $N_c$ large, replacing one of these other
 light quarks with a heavy quark has a negligible effect on this mean color
 field.  Consequently, for large $N_c$, the mass of the light degrees of
 freedom in a baryon containing a single heavy quark is equal to $M_B$, with
 corrections to this relationship of order $N_c^0$.

In the bound state soliton picture $\Lambda_Q$-type bound states arise
 when the spin of the light degrees of freedom of the heavy meson and the spin
 of the nucleon are combined into a spin zero configuration,  and the
 isospin of the heavy meson and the nucleon are combined into an isospin zero
 state. Other baryons (e.g., the $\Delta$) only contribute to bound states with
 higher isospin. The spatial wavefunctions for $\Lambda_Q$-type bound states
 are controlled by the potential [11]
$$	V_\Lambda (\vec x) = V_0 + {1\over 2} \kappa \vec x^2. \eqno (1)$$
Higher powers of $\vec x$ are unimportant for large $N_c$.  Both $V_0$ and
 $\kappa$ are order $N_c^0$ and their values depend on nonperturbative strong
 interaction dynamics.   When the orbital angular momentum of the bound state
 is non-zero, the $\Lambda_Q$-type baryons occur (in the $m_Q \rightarrow
\infty$ limit) in degenerate doublets that
 arise from combining the orbital angular momentum of the bound state with the
 heavy quark spin. The harmonic oscillator potential in eq. (1) gives rise to
an
 infinite tower of $\Lambda_Q$-type baryons with excitation energies
$$\Delta E_{(n_1,n_2,n_3)}^{(Q)}=(n_1+n_2+n_3)\sqrt {\kappa/\mu_Q},  \eqno
 (2)$$
where $\mu_Q$ is the reduced mass
$${1\over \mu_Q}= {1\over m_Q}+{1\over M_B}.   \eqno(3)$$
In eq. (2) $(n_1,n_2,n_3)$ are the quantum numbers that specify the bound
 states when the Schr\"odinger equation is solved by separating variables in
 cartesian coordinates.

For states with the same quantum numbers $(n_1, n_2, n_3)$, but different heavy
quarks, eq. (2) gives
$$\Delta E^{(c)}/\Delta E^{(b)}=\biggl({1+M_B/m_c\over 1+M_B/m_b}\biggr)
^{1\over2} \simeq 1+{1\over 2}\biggl({M_B\over m_c} - {M_B\over m_b}\biggr)
  +...~~~~    .  \eqno(4)$$
Eq. (2) was obtained by solving the Schr\"odinger equation including the
kinetic
 energy of the heavy meson. This corresponds to taking simultaneously the
limits $ m_Q
\rightarrow
 \infty$ and $N_c \rightarrow \infty$ with the ratio $M_B/m_Q$ held fixed
(recall $M_B$ is of order $N_c$). If $m_Q$ was taken to infinity first, then
 effects of order $M_B/m_Q$ are neglected, and heavy quark flavor symmetry
 determines the ratio of excitation energies in eq. (4) to be unity. In the
 large $N_c$ limit the leading corrections to heavy quark symmetry [14] arise
 from
 including the kinetic energy of the heavy meson in the Schr\"odinger equation
for the soliton-heavy meson bound state [11]. This violates heavy quark flavor
 symmetry but leaves heavy quark spin symmetry intact.  Despite the fact that
$M_B/m_c$ is not particularly small the ratio of excitation energies in eq. (4)
differs from unity by less than 20\%.

The excitation energies given in eq. (2) are of order $N_c^{-1/2}$. The first
 excited states have quantum numbers $(1,0,0)$, $(0,1,0)$, $(0,0,1)$. There is
  another basis of quantum numbers $[N,\ell,m]$, $N=n_1+n_2+n_3$, that is also
 useful. Here $\ell$ is the orbital angular momentum of the bound state and $m$
 is the component of the orbital angular momentum along the third
 (spin-quantization) axis. In this basis $N\ge \ell$ and even values of $\ell$
 occur for $N$ even while odd values of $\ell$ occur for $N$ odd. The first
 excited states have $N=1$,
 $\ell =1$, and $m= 0,+1,-1$ giving $s_{\ell}^{\pi_{\ell}}=1^-$ for the spin
 parity of the light degrees of freedom. Combining this with the spin of the
 heavy quark gives a doublet of negative parity states with total spins $3/2$
an $1/2$. For $Q=c$ these states correspond to the observed doublet of excited
 $\Lambda_c$
states. Comparing eq. (2) with the experimental value of the excitation energy
 ($\simeq 340~MeV$) gives $\kappa \simeq (440~MeV)^3$.

In general, $\Lambda_Q$-type states have total spins $s=\ell \pm 1/2$
 formed by combining the spin of the heavy quark with the orbital angular
 momentum $\ell$. We label them by the quantum numbers $\{ N,\ell;s,m\}$, where
 now $m$ is the component of the total spin along the third (i.e.,
 spin-quantization) axis. In this notation the ground state $\Lambda_Q$ baryon
 has quantum numbers $\{ 0,0;1/2,m\}$ and the first excited $\Lambda_Q$ doublet
 contains the states $\{ 1,1;1/2,m\}$ and $\{1,1;3/2,m\}$.

Ref. [11] considered the weak semileptonic decay $\Lambda_b \rightarrow
 \Lambda_c
 e \bar\nu_e$ in the large $N_c$ limit. In this paper we discuss weak
 semileptonic decays of the $\Lambda_b$ to excited $\Lambda_c$ baryons. These
 decay amplitudes are determined by matrix elements of the vector and axial
 vector heavy quark currents. These matrix elements can be calculated using the
bound state soliton picture we have
 outlined above. We find in the rest frame of the initial state, $v=(1,\vec
0)$, that
$$\bigl\langle\Lambda_c^{\{ N,\ell;s^{\prime},m^{\prime}\} }(v^{\prime})|~\bar
 h_{v^{\prime}}^{(c)}\gamma^{\mu}h_v^{(b)}~|\Lambda_b^{\{0,0;1/2,m\}}(v)
\bigr\rangle$$
$$=\delta^{\mu,0}
(\ell,m^{\prime}-m;1/2,m|s^{\prime},m^{\prime})
{\cal F}^{[N,\ell,m^{\prime}-m]}        \eqno(5a)$$
and
$$\bigl\langle\Lambda_c^{\{ N,\ell;s^{\prime},m^{\prime}\} }(v^{\prime})|~\bar
 h_{v^{\prime}}^{(c)}\gamma^{\mu}\gamma_5
h_v^{(b)}~|\Lambda_b^{\{0,0;1/2,m\}}(v)
\bigr\rangle$$
$$=\delta^{\mu,j}\sum_{m^{\prime \prime}}
(\ell,m^{\prime}-m^{\prime \prime};1/2,m^{\prime
\prime}|s^{\prime},m^{\prime})[\chi^{\dagger} (m^{\prime \prime})\sigma^j\chi
 (m)]
{\cal F}^{[N,\ell,m^{\prime}-m^{\prime \prime}]},        \eqno(5b)$$
where ${\cal F}^{[N,\ell,m]}$ is an overlap of momentum space harmonic
 oscillator wave functions
$$	{\cal F}^{[N,\ell,m]} = \int d^3 q \phi_c^{*[N,\ell,m]} (\vec q)
 \phi_b^{[0,0,0]} (\vec q - M_B\vec v^{\prime}). \eqno (6)$$
In eq. (5b) $\chi$ is a two-component Pauli spinor.   The sum over $m''$ in eq.
 (5b) collapses to a single term since $\chi^{\dagger} (m'') \sigma^3 \chi (m)$
 vanishes for $m'' = - m$ and $\chi^{\dagger} (m'') \sigma^{1,2} \chi (m)$
 vanishes for $m'' = m$.  In eq. (6) $\phi_Q^{[N,\ell,m]} (\vec q)$ denotes the
 normalized momentum space harmonic oscillator wave function.  Its dependence
 on the type of heavy quark arises from the dependence of the reduced mass
 $\mu_Q$ on the heavy quark mass.

Eqs. (5) and (6) are valid in the kinematic region $|\vec v'| \lsim {\cal O}
 (N_c^{-3/4})$. For recoil velocities greater than this the overlap ${\cal
 F}^{[N,\ell,m]}$ is very small and terms subdominant in $N_c$ that we have
 neglected may be important [11].  When $v \cdot v' \not=1$ the operator $\bar
h_{v'}^{(c)} \Gamma h_v^{(b)}$ requires renormalization [15].  However, in the
kinematic regime very near zero recoil where eqs. (5) and (6) apply, the
subtraction point dependence of $\bar h_{v'}^{(c)} \Gamma h_v^{(b)}$ is
negligible.

It is easiest to evaluate  ${\cal F}^{[N,\ell,m]}$ in the case where $\vec v'$
 is directed along the 3rd (i.e., spin-quantization) axis.  The expression is
 particularly simple when the limit $m_Q \rightarrow \infty$ is taken first so
 that $\mu_Q = M_B$ independent of heavy quark type.  Then
$$	{\cal F}^{[N,\ell,m]} = \delta^{m,0} {C^{N\ell }\over \sqrt{N!}}
  [M_B^3/\kappa]^{N/4} (v \cdot v' - 1)^{N/2} \exp \left(-{1\over 2}
 [M_B^3/\kappa]^{1/2} (v \cdot v' - 1)\right) \eqno (7)$$
where
$$	C^{N\ell} = \int d^3 q \phi^{*[N,\ell,0]} (\vec q) \phi^{(0,0,N)} (\vec
 q) . \eqno (8)$$
In eq. (7) we used $|\vec v'|^2 = 2 (v \cdot v' - 1)$  which is appropriate for
 the region near zero recoil, $(v \cdot v' -1) \lsim {\cal O} (N_c^{-3/2})$,
 that we are considering.

The ground state $\Lambda_b \rightarrow \Lambda_c$ transition corresponds to
 the case $N = 0,~ \ell = 0$.  Heavy quark spin symmetry implies that in this
 case the matrix elements of heavy quark bilinears have the form [16]
$$	\langle \Lambda_c^{\{0,0;1/2,m'\}} (v')  |\bar h_{v'}^{(c)} \Gamma
 h_v^{(b)}| \Lambda_b^{\{0,0;1/2,m\}} (v)\rangle$$
$$	= \eta (v \cdot v') \bar u (v',  s') \Gamma u(v,s). \eqno (9)$$
Comparing with eqs. (5) and using eqs. (7) and (8) gives [11]
$$ 	\eta (v \cdot v') = \exp \left[- {1\over 2} [M_B^3/\kappa ]^{1/2} (v
 \cdot v' -1)\right]. \eqno (10)$$
Expanding $\eta$ about zero recoil
$$	\eta (v \cdot v') = 1 - \rho^2 (v \cdot v' -1) +... ~, \eqno (11)$$
and comparing with eq. (10) we see that
$$	\rho^2 = {1\over 2} [M_B^3 /\kappa]^{1/2}. \eqno (12)$$

Transition matrix elements to excited $\Lambda_c$ states with $\ell = 1$ are
constrained by
 heavy quark symmetry to have the form [17]
$$	\langle \Lambda^{\{N,1;1/2,m'\}} (v') |\bar h_{v'}^{(c)} \Gamma
 h_v^{(b)} | \Lambda_b^{\{0,0;1/2,m\}} (v) \rangle$$
$$	= {\sigma^{(N)} (v \cdot v')\over \sqrt{3}} \bar u (v', s') \gamma_5
 (\vslash + v \cdot v') \Gamma u(v,s), \eqno (13a)$$
$$	\langle \Lambda_c^{\{N,1;3/2, m'\}} (v')|\bar h_{v'}^{(c)} \Gamma
 h_v^{(b)} | \Lambda_b^{\{0,0;1/2,m\}} (v) \rangle$$
$$	= \sigma^{(N)} (v \cdot v') \bar u_\mu (v',s') v^\mu \Gamma
 u(v,s).\eqno (13b)$$
Comparing these expressions with eqs. (5), (6) and (7) gives
$$	\sigma^{(N)} (v \cdot v') = {C^{N1}\over \sqrt{2(N!)}}
 [M_B^3/\kappa]^{N/4} (v \cdot v' -1)^{(N-1)/2}$$
$$	\cdot \exp \left[- {1\over 2} [M_B^3 /\kappa ]^{1/2} (v \cdot v' -1)\right].
\eqno
 (14)$$
 Note that fractional powers of $(v\cdot v'-1)$ do not occur in eq. (14)
because $N$ must be odd. At zero recoil $\sigma^N (1)$ is zero for $N > 1$
while for the first excited state
$$	\sigma^{(1)}(1) = [M_B^3/4\kappa]^{1/4}, \eqno (15)$$
using $C^{11} = 1$. Equations (14) and (15) are the main results of this paper.
 For simplicity we derived our expressions for the Isgur--Wise functions
 $\sigma^{(N)}(v\cdot v')$ by taking the limit $m_Q \rightarrow \infty$
 followed by
the limit $N_c \rightarrow \infty$. However, eqs. (5) and (6) can be used to
 include corrections to the heavy quark limit to all orders in  $M_B/m_Q$. As
we
 have noted, these corrections do not violate heavy quark spin symmetry.
 Therefore the form of the matrix elements given in eqs. (13) still holds, but
the functions  $\sigma^{(N)}(v\cdot v')$ become dependent on the heavy quark
 masses.

The fact that $\sigma^{(N)}(1)$ is zero for $N>1$ means that in the large $N_c$
 limit the Bjorken sum rule [17,18] for the slope $\rho^2$ of the Isgur--Wise
 function $\eta (v\cdot v')$ and Voloshin sum rule [19] for the mass of the
light degrees of freedom are saturated by the first doublet of excited
 $\Lambda_c$ states.  The Bjorken sum rule for the slope of the Isgur--Wise
 function $\eta(v\cdot v')$ is [17]
$$	\rho^2 = \sum_N |\sigma^{(N)} (1)|^2, \eqno (16)$$
while the Voloshin sum rule for the mass of the light degrees of freedom, $\bar
 \Lambda = M_{\Lambda_{Q}} - m_Q$, reads
$$	\bar \Lambda = \sum_N 2 \Delta E_N  |\sigma^{(N)} (1)|^2. \eqno (17)$$
(Eq. (17) is a generalization of the heavy meson result derived in Ref. [19] to
 the case of heavy baryons.) Using eq. (15) and our explicit expressions for
 $\Delta E_N,~ \rho^2$ and $\bar \Lambda = M_B$, it is straightforward to
 verify that in the large $N_c$ limit these two sum rules hold.

There are other properties of excited heavy baryons that can be examined in the
 large $N_c$ limit. For example, at the leading order in chiral perturbation
 theory [20], the strong couplings of the ground state $\Lambda_Q$  to
 $\Sigma_Q\pi$ or
$\Sigma_Q^* \pi$ are of order $N_c^{1/2}$ and can be related to the
 pion-nucleon coupling [9]. However, because of the orthogonality of the
 harmonic oscillator wave-functions the analogous couplings for excited
 $\Lambda_Q$ states [21] are only of order $N_c^{-1/2}$.
\vfil\eject

\centerline {\bf References}
\smallskip
\item{1.}   H. Albrecht, et al., (ARGUS Collaboration), Phys. Lett., {\bf B317}
 (1993) 227; D. Acosta, et al., (CLEO Collaboration), CLEO CONF 93-7;
Contributed
 to the International Symposium on Lepton and Photon Interactions, Ithaca,
 (1993); M. Battle, et al., (CLEO Collaboration), CLEO CONF 93-32; Contributed
to
 the International Symposium on Lepton and Photon Interactions, Ithaca, (1993);
 P. L. Frabetti, et al., (E687 Collaboration), FERMILAB-Pub-93-32-E (1993).

\item{2.} L. A. Copley, N. Isgur and G. Karl, Phys. Rev., {\bf D20} (1979)
 768.

\item{3.} G. `t Hooft, Nucl. Phys., {\bf B72} (1974) 461; {\bf B75} (1974)
 461.

\item{4.} E. Witten, Nucl. Phys., {\bf B160} (1979) 57.

\item{5.} T.H.R. Skyrme, Proc. Roy. Soc., {\bf A260} (1961) 127; E. Witten,
 Nucl. Phys., {\bf B223} (1983) 433; G.S. Adkins, et al., Nucl. Phys., {\bf
B228}  (1983) 552.

\item{6.} C.G. Callan and I. Klebanov, Nucl. Phys., {\bf B262} (1985) 365;
 Phys. Lett., {\bf B202} (1988) 269.

\item{7.} M. Rho, et al., Phys. Lett., {\bf B251} (1990) 597; Z. Phys., {\bf
 A341} (1992) 343; D.O. Riska and N.N. Scoccola, Phys. Lett., {\bf B265} (1991)
 188; Y. Oh, et al., Nucl Phys., {\bf A534} (1991) 493.

\item{8.} E. Jenkins, et al., Nucl. Phys., {\bf B396} (1993) 27.

\item{9.} Z. Guralnik, et al., Nucl. Phys., {\bf B390} (1993) 474.

\item{10.} E. Jenkins and A.V. Manohar, Phys. Lett., {\bf B294} (1992) 273.

\item{11.} E. Jenkins, et al., Nucl. Phys., {\bf B396} (1993) 38.

\item{12.} D.P. Min, et al., SNUTP-92-78 (1992); M.A. Nowak, et al., Phys.
 Lett., {\bf B303} (1993) 130; K.S. Gupta, et al., Phys. Rev., {\bf D47} (1993)
 4835; J. Schechter and  A. Subbaraman, Phys. Rev., {\bf D48} (1993) 332;  Y.
Oh, et al., SNUTP-93/80 (1993).

\item{13.} E. Jenkins, Phys. Lett., {\bf B315} (1993) 431; {\bf B315} (1993)
447.

\item{14.}  N. Isgur and M.B. Wise, Phys Lett. {\bf B232} (1989) 113; {\bf
 B237} (1990) 527.

\item{15.} A. Falk, et al., Nucl. Phys., {\bf B343} (1990) 1.

\item{16.} N. Isgur and M.B. Wise, Nucl. Phys., {\bf B348} (1991) 278; H.
 Georgi, Nucl. Phys., {\bf B348} (1991) 293; T. Mannel, et al., Nucl. Phys.,
{\bf B355} (1991) 38; F. Hussain, et al., Z. Phys. {\bf C51} (1991) 321.

\item{17.} N. Isgur, et al., Phys. Lett. {\bf B254} (1991) 215.

\item{18.} J.D. Bjorken, SLAC-PUB-5728, invited talk presented at Recontre
 de Physique de la Vallee D'Acoste, La Thuile, Italy (1990).

\item{19.} M.B. Voloshin, Phys. Rev., {\bf D46} (1992) 3062.

\item{20.} P. Cho, Phys. Lett., {\bf B285} (1992) 145.

\item{21.} P. Cho, CALT-68-1912 (1994).

\bye